\documentclass[fleqn,twoside]{article}
\usepackage{espcrc2}

\usepackage{graphicx}
\usepackage[figuresright]{rotating}

\newcommand{\AmS}{{\protect\the\textfont2
  A\kern-.1667em\lower.5ex\hbox{M}\kern-.125emS}}

\title{High-Energy Spectra From Black-Hole Candidates}

\author{T. Belloni\address[OAB]{INAF-Osservatorio Astronomico di Brera,
	Via E. Bianchi 46, I-23807 Merate, Italy}}%
       
\begin{document}

\begin{abstract}

Our knowledge of the high-energy emission spectra from black-hole candidates 
has increased enormously with missions like BeppoSAX and RossiXTE, thanks to 
their broad-band window. I present the main and most solid points of this 
current knowledge, both in terms of detailed spectral modeling of X-ray
broad-band spectra, and more 
generally in relation to the ``black-hole state paradigm", which links spectral 
and timing properties of accretion disks. Finally, through the discussion
of recent results on two systems, XTE J1650-500 and GX 339-4, 
I present evidence for a new view of the state paradigm.

\vspace{1pc}
\end{abstract}

\maketitle

\section{Introduction}

The launch of RossiXTE and BeppoSAX have opened a new era for our
understanding of high-energy spectra from Black-Hole Candidates (BHC).
On the one hand, the large number of pointings allowed by the flexibility
of RXTE, both for transient and persistent systems, has allowed a 
thorough coverage of spectral (and timing) properties of these systems,
in particular with respect to their state transitions. On the other hand,
the broad-band capabilities of both satellites have given us high
signal-to-noise ratio spectra over almost three decades in energy. This 
made it possible to study in detail the overall spectral distribution
without leaving out any important component.
In the following, after a brief description of the current status
of the state-classification, I will first discuss 
some general issues about broad-band
spectra of BHC, based on a few examples. Then, I will present detailed
results on two systems observed with RXTE, XTE J1650-500 and GX 339-4, 
concluding with a new view of BH canonical states.

\section{``Canonical states''}

After a long history of changes, additions and removals, the author's current
picture of BH states is the following (see \cite{tl95,vdk95,men97,hom01}). 
\begin{itemize}

\item {\it LS:} Low/hard state.
        The energy spectrum is dominated by a power-law-like 
	component, with a photon index $\sim$1.6, which shows a clear
	high-energy cutoff at around $\sim$100 keV. An additional weak
	very soft component, probably associated to the thermal disk,
	might be observable below 1 keV if the interstellar absorption is not
	too high.

\item {\it HS:} High/soft state.
	The energy spectrum is dominated by a thermal component, usually
	modeled with a disk-blackbody component with a temperature of
	1-2 keV. A weak power-law component is present,with a steeper photon
	index. No apparent high-energy cutoff is observed in high
	signal-to-noise spectra.

\item {\it VHS/IS:} Very High/Intermediate state.
	The energy spectrum features both a thermal component similar to 
	that observed in the HS and a hard power-law like component
	with photon index $\sim$2.5. The relative contribution of these
	two components to the 1-10 keV flux can vary between 10 and 90\%.

\end{itemize}

In the above I did not discuss the presence of additional components as
Compton reflection or iron lines/edges as I wanted to limit myself to 
he main components at play.
Notice that I also did not mention variability features, which are important
in the definition of states. This is intended, in order to focus the
attention on to the spectral properties, although indeed in some case
the knowledge of timing properties would be necessary to assess the current
state of a source.

After the results obtained on  XTE J1550-564 \cite{hom01}, it is now
clear that changes in the mass accretion rate are not sufficient to 
explain all state transitions, and that there must be a second parameter
at play. The nature of this second parameter is at present unknown.

\section{Broad-band X-ray spectra}

The availability of better instruments and missions has put in our hands
extraordinary tools to characterize broad-band spectra of BHC. In 
particular, the quality of BeppoSAX spectra obtained combining all four
narrow-field instruments on board (LECS, MECS, HPGSPC and PDS) was
extraordinary. The
increase in quality however led to an increase in complication of the
models used for spectral fitting. This makes it necessary to be much more
careful in the modeling and interpretation, keeping also in mind that even
the most detailed model cannot be but a simplification of reality.
It is important to keep always in mind that X-ray spectral fitting at 
such energy resolutions yields an output that is strongly dependent on the
assumptions made on the spectral model.

I show here three important examples.

\subsection{XTE J1118+480}

Probably the best source for a broad-band energy study was XTE J1118+480, 
as its low interstellar absorption allowed to detect flux down to well
below 1 keV (see \cite{mcc01,fro01}). This source was observed only
in the Low/Hard state.
A large multi-wavelength campaign led to a detailed study of the 
Spectral Energy Distribution (SED), covering more than five
orders of magnitude in energy \cite{hyn00,mcc01,esi01,mar01}. 

\begin{figure}[htb]
\includegraphics[width=7.5 cm]{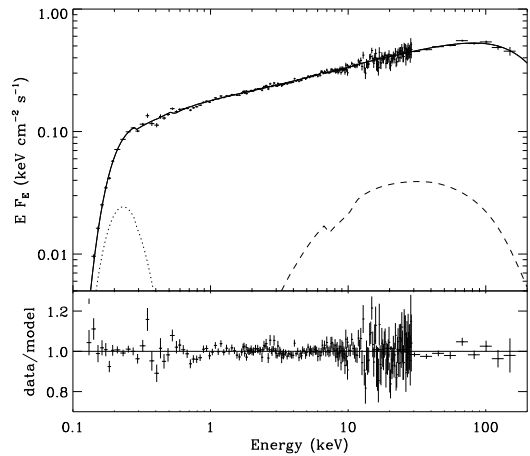}
\caption{Deconvolved spectrum corresponding to the complex fit described in
text for XTE J1118+480 (from \cite{fro01}).}
\label{fig1}
\end{figure}

The broad-band X-ray
spectrum was studied from BeppoSAX observations \cite{fro01,fro02}.
The 0.1-200 keV spectrum was observed to be rather complex. The preferred
best-fitting model consisted of a soft multi-temperature black body
with inner temperature between 35 and 52 eV, observable only because of
the low value for the interstellar absorption 
($\sim 1.3\times 10^{20}$cm$^{-2}$), plus a thermal Comptonization model
with electron temperature T$_e\sim$85 keV and optical depth around unity.
In addition, a weak Compton reflection component was included (with
covering fraction R$\sim$0.2).
The latter component indicated  a low metallicity value ($Z/Z_\odot\sim 0.13$),
consistent with the high galactic latitude of the system. 
This is a very complex model, involving an actual fit to the metallicity of the
gas, which needs to be taken cautiously because of the number of theoretical
assumptions needed to justify it. Indeed the first of the four spectra
analyzed by \cite{fro02} can be fitted also with a simpler model consisting
of a multi-T black body and a power law with a high-energy cutoff. The
magnitude of the deviations introduced by the Thermal Comptonization and
Compton reflection models can be appreciated in Fig. \ref{fig1}, where the best
fit to the complex model is shown in a deconvolved fashion. Nevertheless,
the possibility of such a detailed fit from simultaneous data opens new
possibilities for physical modeling of these spectra.

\subsection{Cygnus X-1}

\begin{figure}[htb]
\includegraphics[width=7 cm]{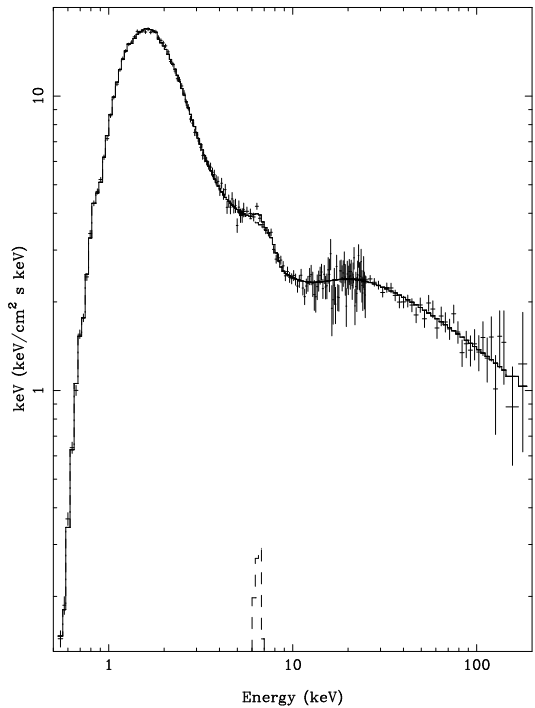}
\caption{Deconvolved spectrum corresponding to the soft-state BeppoSAX
observation of Cygnus X-1 (from \cite{frocy}).}
\label{fig2}
\end{figure}

It is difficult not to discuss the first and most famous BHC, Cyg X-1. 
It was observed by BeppoSAX both during its low/hard (LS) and high/soft (HS) 
states \cite{frocy}. Again, the spectra are interpreted with a very complex
model, rather similar in nature to the one discussed above. Detailed
spectral modeling of a RXTE observation to Cyg X-1 in the hard state
has been also reported, with simpler assumptions on the models
(see \cite{dov98}).
What is interesting here is the comparison between the two states. The hard
Comptonizing component is fitted to a purely thermal model in the case
of the hard state, and with a hybrid model for the HS (see Fig.
\ref{fig2}). Once again, the characterization of the hard component,
thanks to the broad-band spectrum, goes into considerabl detail.

\subsection{Thermal vs. non-thermal}

\begin{figure}[htb]
\includegraphics[width=7.5 cm]{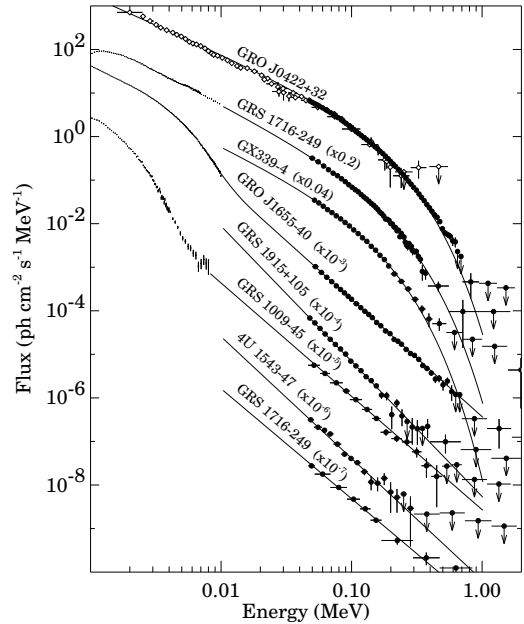}
\caption{Deconvolved 1 keV-1 MeV spectra of a number of BHC observed by
OSSE and other instruments (from \cite{gro98}).}
\label{fig3}
\end{figure}

The different nature of the hard component in different states had been
discovered earlier (see e.g. \cite{gil95}). 
An important survey of a number of BHC made with OSSE
showed clearly that the presence of an ultrasoft component in the 1-10 keV
range was associated to the lack of a detectable high-energy cutoff in the
high-energy spectra up to 1 MeV, indicating a probable non-thermal origin
\cite{gro98}. A collection of spectra can be seen in Fig. \ref{fig3}.

It is important to notice that it is not clear whether the spectra with
no high-energy cutoff correspond to the soft/high state or to the
very high/intermediate state (see \cite{hom01}). It is therefore possible
that also VHS/IS high-energy spectra have a non-thermal origin.

\subsection{GRS 1915+105}

\begin{figure}[htb]
\includegraphics[width=7 cm]{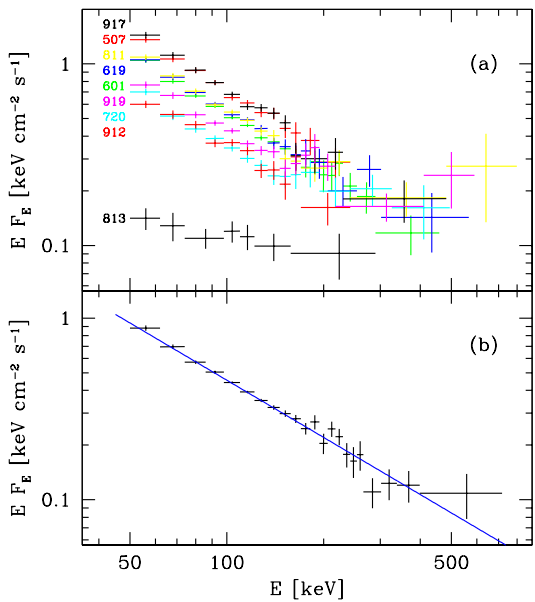}
\caption{Top panel: nine OSSE spectra of GRS 1915+105 deconvolved with 
a power law. Bottom panel: average spectrum (from \cite{zdz01}).}
\label{fig4}
\end{figure}

This source shows the most complex phenomenology ever observed in X-ray
binaries (see \cite{bel00}). However, there is evidence that, at least
most of the time, the source is in something similar to the VHS of
more standard systems \cite{bel98,rei03}. A combination of RXTE and OSSE
data has shown that both in its B and C states, the hard component of 
GRS 1915+105 does not show any high-energy cutoff up to 0.9 MeV \cite{zdz01}.
The OSSE spectra can be seen in Fig. \ref{fig4}.

\subsection{Which component?}

From the data shown above, it would seem that the component observed in the VHS
is associated to that of the soft state and not to the thermal spectrum
observed in the hard state. However, this is in contradiction with the
properties of the timing variability, which shows properties gradually
changing from the hard to the very high state (see e.g. \cite{bel03}).
In other words, the timing results associate the VHS hard component to that
seen in the HS, while spectra considerations connect it with that seen in 
the LS. This inconsistency needs to be resolved in order to present a 
clear view for the development of theoretical models. Integral is
probably the mission that will give definite answers on the spectral side
(see \cite{rod03}).

\section{BH transients: two recent examples}

The best way to obtain information about states and state-transitions
is to observe transient systems with dense campaigns that allow to monitor
the timing/spectral properties of these systems for a whole outburst.
In this way, one obtains a more coherent view of the properties of
accretion in these systems, a more promising approach than the extremely
detailed spectral fitting to a small number of observations are shown
above.
I outline here the basic results obtained on two systems, based on a large
number of RXTE observations \cite{ros02,ros03,bel03,nesqp,nes03,hom04}.

\subsection{XTE J1650-500}

This transient was discovered by the ASM/RXTE in September 2001, when
an outburst lasting about fifteen months started. Follow-up observations showed
a hard spectrum, strong variability, low and high frequency QPOs, and
a relativistic skewed iron line suggesting the presence of a Kerr black
hole \cite{mak01,rev01,wij01,hom03,mil02}. The final portion of the 
outburst was studied by \cite{kal03}.

\begin{figure}[htb]
\includegraphics[width=7 cm]{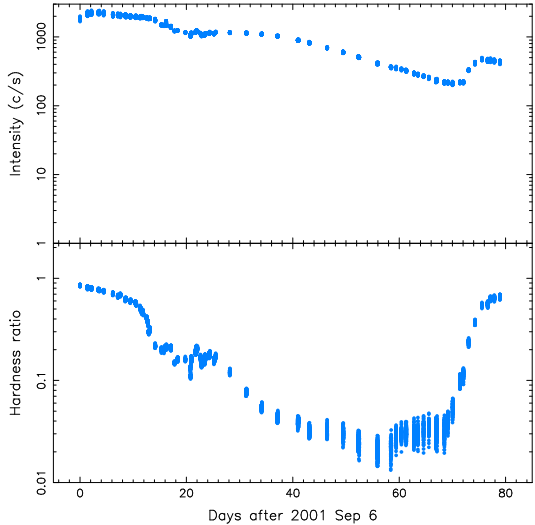}
\caption{Top panel: PCA slight curve (PCU2 only) for the XTE J1650-500 
RXTE campaign. Each point is accumulated over 16 s.
Bottom panel: corresponding hardness curve (see text). (from \cite{ros03}).}
\label{xtelicu}
\end{figure}

For the bright part of the outburst, more than one observation per day was
performed with RXTE, yielding the most complete coverage of such a transient.
The full spectral and timing analysis is presented in \cite{ros02,ros03}.
Here I show the basic evolution of the outburst. The PCA light curve
and the corresponding X-ray hardness evolution (the hardness ratio
is defined as the ratio of the counts in the 4.5-7.9 keV band to those in the
2.5-4.6 keV band) is shown in Fig. \ref{xtelicu}. 

\begin{figure}[htb]
\includegraphics[width=7 cm]{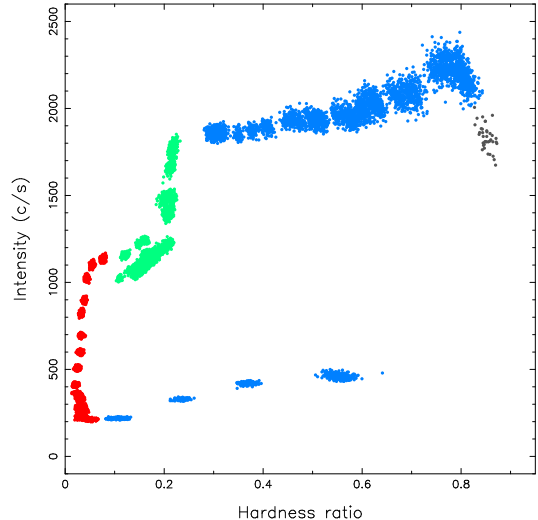}
\caption{Hardness-Intensity diagram for XTE J1650-500 (PCU2 only)
corresponding to the
points shown in Fig. \ref{xtelicu}. (from \cite{ros03}).}
\label{xtehid}
\end{figure}

\begin{figure}[htb]
\includegraphics[width=7 cm]{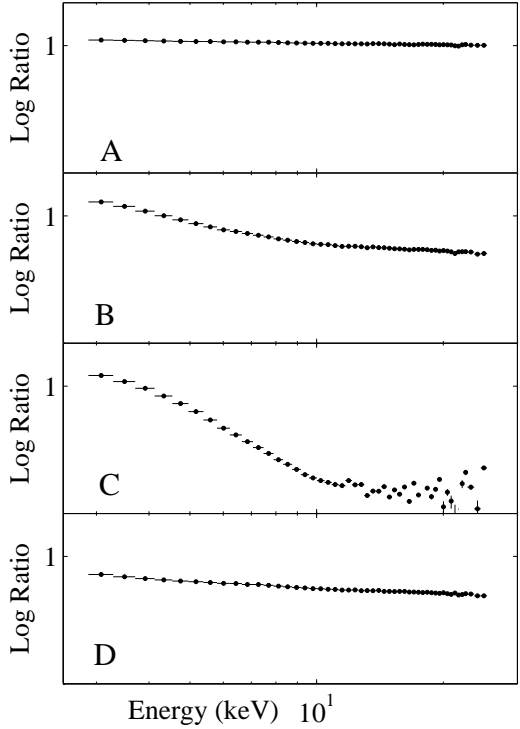}
\caption{Ratio of the energy spectrum of four representative 
observations to that of the first observation of the campaign of XTE
J1650-500. Panel A: observation on the upper horizontal track in 
Fig. \ref{xtehid}; panel B: observation on the stable interval (see text);
panel C: observation from the soft track; panel D: observation from the
lower horizontal track.
}
\label{ratios}
\end{figure}

The density of the coverage is evident, as is the large
range of spectral variations. A better view of these variations can be
obtained by a Hardness-Intensity diagram (HID) (see Fig. \ref{xtehid}).
The source starts at the upper right corner, already at a bright level and
hard, then moves left softening gradually. After a drop in count rate, it
stabilizes for about ten days on to a small diagonally-shaped region of the HID.
After this, it follows a vertical line as it becomes steadily fainter and
slightly softer, to enter another horizontal hardening line. Not shown here,
in the last part of the outburst, the flux decreases considerably, while the
hardness remains approximately at the values found at the beginning of the
outburst.
Detailed spectral analysis in terms of the two main components shows that 
on the two horizontal tracks, both a soft and a hard component are present,
with the hard one dominating the flux. During the stable interval, 
it is the soft
component that dominates. Finally, along the vertical track, 
the hard component is
very weak and almost all the flux comes from the soft component.
This spectral evolution can be seen  in Fig. \ref{ratios}. The four panels
show for four representative PCA observations the ratio of the energy
spectrum over that of the first observation, where the spectrum was
dominated by a flat power law. Panel {\it A} corresponds to an observation
from the top track in Fig. \ref{xtehid}: here the ratio indicates a
power-law spectrum similar to the one at the beginning.
Panel {\it B} corresponds to an observation
from the stable interval: the presence of a soft component is evident, 
as well as a steepening of the power law.
Panel {\it C},  an observation from the soft branch, shows a spectrum
dominated by the soft component, and 
finally panel {\it D}, from the low horizontal track, shows again a power law,
slightly steeper than panel {\it A}

We can identify the horizontal tracks and the stable section as VHS, and
the vertical track as HS, identification confirmed by the timing analysis
\cite{ros03}. The LS is not observed here at the beginning, as most of the
rising part of the outburst was not observed, while it was observed
at the end, after the observations shown here \cite{kal03}.

\subsection{GX 339-4}

GX 339-4 is historically an important system, as it was the only persistent
BHC that showed all spectral/timing states \cite{miy91,ebi94,bel97,men97,bel99}.
In May 1999, it entered a period of quiescence that lasted until 2002
\cite{kon00,cor03}. In March 2002, a new long outburst began and the
source was observed regularly by RXTE for more than a year 
\cite{bel03,nesqp,nes03}.

\begin{figure}[htb]
\includegraphics[width=7 cm]{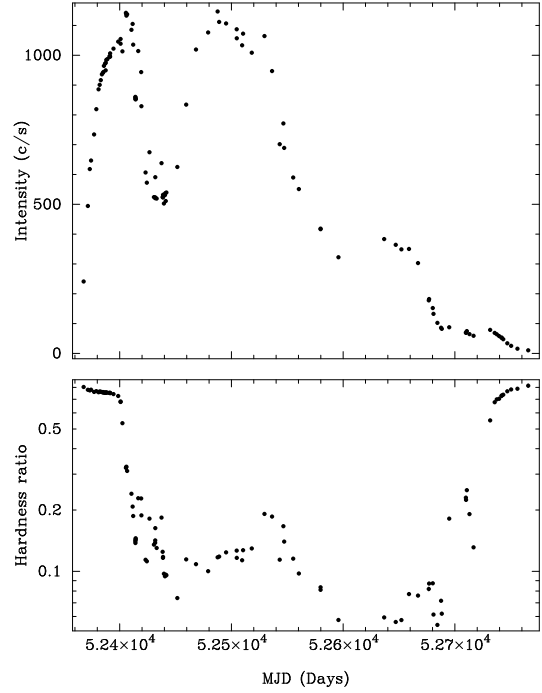}
\caption{Top panel: PCA slight curve (PCU2 only) for the GX 339-4
RXTE campaign. Each point  correspond to one observation.
Bottom panel: corresponding hardness curve (see text). (from \cite{nes03}).}
\label{339licu}
\end{figure}

\begin{figure}[htb]
\includegraphics[width=7 cm]{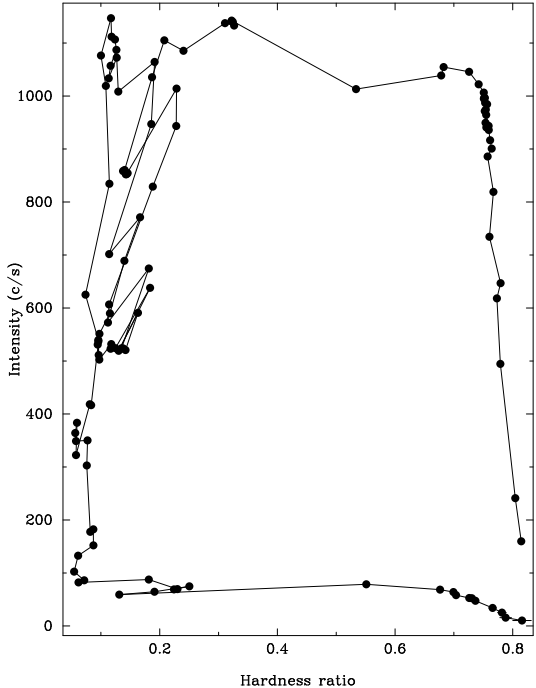}
\caption{Hardness-Intensity diagram for GX 339-4 (PCU2 only)
corresponding to the
points shown in Fig. \ref{339licu} (from \cite{nes03}).}
\label{339hid}
\end{figure}

Figure \ref{339licu} shows the PCA light and hardness curves for GX 339-4,
where hardness is now the ratio of the counts in the 4.5-8.6 keV band 
to those in the 2.0-4.6 keV band
(from \cite{nes03}). Large hardness variations are visible. Once again,
the spectral changes can be seen better in a HID (Fig. \ref{339hid}).
Here the coverage of the outburst started earlier, and the source is 
observed to start from the bottom right and move upwards along a vertical
track, consistent with a constant spectrum and flux variations. After this
initial phase, in its general trend the HID is similar to that of XTE J1650-500.
First a horizontal track at high count rates shows the source softening. Then
there are a few observations spent in a region of lower count rate, followed
by a movement onto the soft vertical branch. At the end of the outburst,
a low-rate horizontal track is followed to the right, to go back to the initial
hardness. 

\section{Another look at the canonical states}

\begin{figure}[htb]
\includegraphics[width=7 cm]{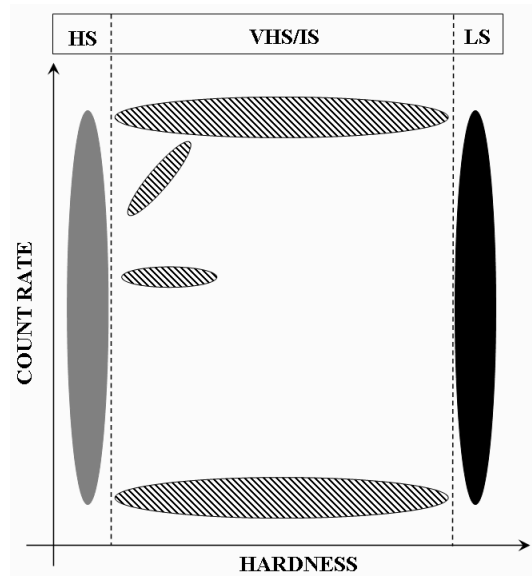}
\caption{Schematic Hardness-Intensity diagram for a generic BHC.
}
\label{hid}
\end{figure}

From the outburst behavior of the two sources examined above, with the
excellent coverage provided by RXTE, we can
take a fresh look at the canonical states from the spectral point of
view. In the following, I present a short summary of work that will
be presented in \cite{hom04}.
A schematic HID is shown in Fig. \ref{hid}. The three states
are clearly identified in the diagram:

\begin{itemize}

\item {\it LS:} it is clearly associated only to the beginning 
	and the end phases
	of the outburst, i.e. to lower accretion rate values, and
	never observed mixed with the other states. The LS is
	represented by the black track on the right side of the 
	diagram. This track extends down to very low count rates,
	spanning three orders of magnitude in the case of GX 339-4
	\cite{nes03}.

\item {\it HS:} associated to the central section of the outburst, it
	is represented by the gray track on the extreme left, where
	the soft thermal component completely dominates the spectrum.

\item {\it VHS/IS:} here defined as the central region of the HID, in 
	between the thin LS and HS vertical tracks. There is clearly
        more than one different instance of VHS/IS, which is reflected
	by the fact that the definition of VHS/IS which can be found
	in the literature is sometimes rather vague.

\end{itemize}

\begin{figure}[htb]
\includegraphics[width=7 cm]{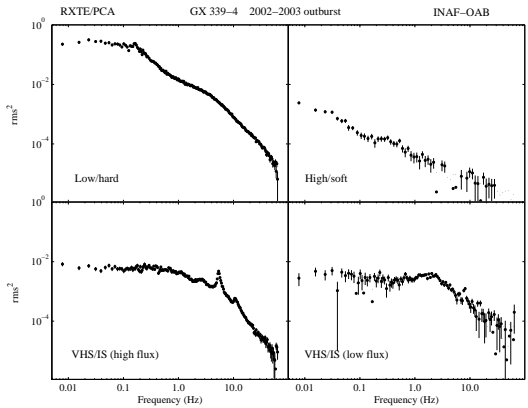}
\caption{Four characteristic power density spectra for GX 339-4 as
	observed by RXTE \cite{nes03}. The upper two panels show
	examples from the HID hard (LS: left) and soft branch (HS:
	right).
	The bottom panels refer to two observations in the VHS/IS, on
	the high-flux (left) and low-flux (right) branch.
}
\label{pds}
\end{figure}

Although I tried to limit myself to spectral information, it is not
possible to leave variability information aside, as it is fundamental
in assessing source states. In Fig. \ref{pds} I show four 
representative Power Density Spectra (PDS) from observations of
GX 339-4. They correspond to four observations from the four tracks
shown in Fig. \ref{339hid}. The top two PDS, from the right and left
vertical tracks, are typical of LS and
HS and can be compared to those from \cite{bel99}. The bottom two,
from the top and bottom horizontal tracks, are typical of VHS and IS,
as also observed in the past from the same source \cite{bel97,men97}.

With this characterization, the VHS/IS, for which Intermediate State is
clearly a better denomination, is configured as a transitional state, 
most notably between LS and HS (or viceversa), but also as a ``transient''
state which can appear for short times in the middle of an outburst.
Whether these three instances of what we call IS are the same physical
state is not clear, but needs to be investigated through theoretical models.

\section{Conclusions}

In the first part of the paper, I presented recent results on the broad-band
spectral distribution of BHC, made possible by the exceptional combination
of instruments on board RossiXTE and especially BeppoSAX. We are now in 
the condition to test complex physical models to the data, although the
limitations of the instruments must be kept into account.
The different nature (thermal/non-thermal) of the hard component was examined.
In particular, I showed how timing and spectral information available
are not consistent in associating the hard component observed in the IS
to either that characteristic of the LS or to that of the HS.

The second part was dedicated to presenting basic spectral results from 
recent RXTE observational campaigns of XTE J1650-500 and GX 339-4. I 
presented in particular Hardness-Intensity diagrams  for the two systems
and compared them. From these results, supported by extensive timing analysis
only briefly shown here, I present a new 
phenomenological view of spectral/timing
states of BHC, clarifying some of the existing confusion between different
flavors of VHS/IS and identifying a partial dependence of state transitions
on variations in the mass accretion rate.

\section{Acknoweledgements}

I thank Jeroen Homan, Sabrina Rossi and Elisa Nespoli, who provided the
results of their analysis of GX 339-4 and XTE J1650-500.

\end{document}